\documentclass[11pt,a4paper]{emulateapj}

\usepackage{natbib}

\newcommand{\teff}{${T}_{\mathrm{eff}}$}
\newcommand{\logg}{$\log{g}$}
\newcommand{\msun}{$M_{\odot}$}
\newcommand{\rsun}{$R_{\odot}$}

\shorttitle{Two New Tidally Distorted WDs}
\shortauthors{Hermes et al.}

\begin{document}

\title{TWO NEW TIDALLY DISTORTED WHITE DWARFS}
\slugcomment{To appear in the Astrophysical Journal}
\author{J. J. Hermes\altaffilmark{1,2}, Mukremin Kilic\altaffilmark{3}, Warren R. Brown\altaffilmark{4}, M. H. Montgomery\altaffilmark{1,2}, and D. E. Winget\altaffilmark{1,2}}

\altaffiltext{1}{Department of Astronomy, University of Texas at Austin, Austin, TX 78712, USA}
\altaffiltext{2}{McDonald Observatory, Fort Davis, TX 79734, USA}
\altaffiltext{3}{Homer L. Dodge Department of Physics and Astronomy, University of Oklahoma, 440 W. Brooks St., Norman, OK 73019, USA}
\altaffiltext{4}{Smithsonian Astrophysical Observatory, 60 Garden St, Cambridge, MA 02138, USA}

\email{jjhermes@astro.as.utexas.edu}


\begin{abstract}

We identify two new tidally distorted white dwarfs (WDs), SDSS J174140.49+652638.7 and J211921.96$-$001825.8 (hereafter J1741 and J2119). Both stars are extremely low-mass (ELM, $\leq0.2 M_{\odot}$) WDs in short-period, detached binary systems. High-speed photometric observations obtained at the McDonald Observatory reveal ellipsoidal variations and Doppler beaming in both systems; J1741, with a minimum companion mass of 1.1 \msun, has one of the strongest Doppler beaming signals ever observed in a binary system (0.59$\pm$0.06\% amplitude). We use the observed ellipsoidal variations to constrain the radius of each WD. For J1741, the star's radius must exceed 0.074 \rsun. For J2119, the radius exceeds 0.10 \rsun. These indirect radius measurements are comparable to the radius measurements for the bloated WD companions to A-stars found by the {\em Kepler} spacecraft, and they constitute some of the largest radii inferred for any WD. Surprisingly, J1741 also appears to show a 0.23$\pm$0.06\% reflection effect, and we discuss possible sources for this excess heating. Both J1741 and J2119 are strong gravitational wave sources, and the time-of-minimum of the ellipsoidal variations can be used to detect the orbital period decay. This may be possible on a timescale of a decade or less.

\end{abstract}



\section{Introduction}

White dwarfs (WDs) are the remnant, degenerate cores of stars. About half of the stars in the galaxy go through stellar evolution as single stars and end up as typical Earth-sized 0.6 \msun\ C/O-core WDs. The remaining half evolve in binary systems, including short-period systems where stable or unstable mass transfer can take place in the late stages of their stellar evolution. Many short-period systems go through one or two common-envelope phases and evolve into systems containing low-mass WDs. Indeed, radial velocity surveys of low-mass, He-core WDs ($M<0.45$\msun) indicate that most form in binary systems \citep{Marsh95,Brown11a}. The lower mass WDs are predicted to be larger in radius than their more massive counterparts. However, observational data for radius measurements for low-mass WDs are scarce.

Eclipsing binaries are important laboratories for constraining the mass-radius relation for WDs, but only a handful of direct radius measurements for low-mass, He-core WDs exist. Observations using the {\em Kepler} spacecraft have so far found four A-stars with bloated $\sim0.2-0.4$\msun\ WD companions. The radius measurements for these four WDs range from 0.04 to 0.15 \rsun\ \citep{vankerkwijk10,Carter11,Breton11}. However, due to their extreme environments, all four of these WDs are predicted to be larger than normal. Hence, they provide only an upper limit for the mass-radius relation for low-mass WDs. In addition, there have been four eclipsing low-mass WD systems detected from the ground \citep{Steinfadt10a,Parsons11,Vennes11,BrownJ0651}. \citet{maxted11} recently identified a 0.23\msun\ stripped core of a red giant star, a pre-He WD. They measure a radius of 0.33 \rsun\ from the eclipse observations, providing an estimate for the radii of the extremely low mass (ELM) WD progenitors.

Tidally distorted WDs in (non-)eclipsing systems also provide reliable constraints on WD radii. The amplitude of the ellipsoidal variations is roughly $\delta f_{ELV} \approx (m_2/m_1)(r_1/a)^3$, where $a$ is the orbital semi-major axis and $r_1$ is the radius of the primary \citep[e.g.,][]{Mazeh10}. Hence, this effect is most important for short-period systems with extreme mass ratios and more importantly with relatively large (radius) WDs, i.e., systems containing ELM WDs. Not surprisingly, the first and second direct detection of tidally distorted WDs occurred in the $P<1$ h orbital period ELM WD systems J0106 \citep{KilicJ0106} and J0651 \citep{BrownJ0651}. J0651, a 12.75-minute orbital period detached eclipsing double WD system, illustrates the wealth of photometric information that can be present in an ELM WD system: there are primary and secondary eclipses, ellipsoidal variations, and a strong Doppler beaming signal.

The ELM Survey \citep{BrownELMi,KilicELMii,BrownELMiii} continues to discover a large number of short-period systems containing low-mass WDs. This survey has been quite successful, so far uncovering two dozen double-degenerate binaries with $<10$ Gyr merger times \citep{KilicELMiv} and the only known detached binary WDs with $<$1 h orbital periods: two 39-minute orbital period binaries, J0106 and J1630 \citep{KilicJ0106,KilicJ1630}, and the 12.75-minute system, J0651 \citep{BrownJ0651}. Based on the probability of eclipses and other photometric variability at orbital period timescales, we have established a follow-up program to observe the shortest-period binaries found in the ELM Survey. These time-series photometric observations have been carried out at the McDonald Observatory using the Argos instrument, a frame-transfer CCD mounted at the prime focus of the 2.1m Otto Struve telescope \citep{Nather04}.

Observations of tidal distortions in these short-period systems can provide important constraints on the physical parameters of the binary. Any mis-alignment between the WD spin period and the binary orbital period can be measured using the ellipsoidal variations; a mis-alignment will result in tidal heating of the ELM WD \citep[e.g.,][]{Piro11,Fuller11}. If the radius of the WD is known, ellipsoidal variations can also be used to constrain the mass of the unseen companion and the inclination angle. This has strong implications for understanding the future evolution of these systems, which depends on the mass ratio of the two stars \citep{Marsh04}.

Here we take advantage of the latest discoveries from the ELM Survey. J1741 and J2119 were recently identified as 1.5-2 hr orbital period systems containing two of the lowest surface gravity WDs currently known ($\log{g}=5.2-5.4$; \citealt{BrownELMi,BrownELMiii}). We report the detection of ellipsoidal variations and Doppler beaming signals in both stars. We discuss our use of the Doppler beaming signal to confirm the radial velocity variations previously observed in J1741, and use the ellipsoidal variations to constrain the radii of the tidally distorted ELM WDs. We discuss our observations and analysis of J1741 and J2119 in Sections 2 and 3, respectively. We summarize our results and the mass-radius relation for low-mass WDs in Section 4.



\section{SDSS J174140.49+652638.7}

J1741 is one of the lowest surface gravity WDs known \citep{BrownELMiii}. It has $T_{\rm eff}=9{,}790 \pm 240$ K, $\log{g}=5.19\pm0.06$, and $M \sim$ 0.16 \msun. J1741 has a $1.4666\pm0.0001$ h orbital period and a mass function of $0.830\pm0.018$ \msun. Given the spectroscopically determined mass of the primary, the minimum mass of the unseen companion is 1.09 \msun, and there is a 57\% chance that the inclination is such that its companion is more massive than 1.4 \msun.

\subsection{The Light Curve}

Photometric observations of J1741 were carried out at the McDonald Observatory over two nights in 2011 May and one night in 2011 September, for a total of more than 9.5 hr of coverage. Exposure times for this $g=18.4$ mag WD ranged from 15 to 20 s, depending on the seeing and sky transparency. Observations were obtained through a 1mm BG40 filter to reduce sky noise. 

We performed weighted aperture photometry on the calibrated frames using the external IRAF package $\textit{ccd\_hsp}$ written by Antonio Kanaan. We divided the sky-subtracted light curves by at least two brighter comparison stars in the field to allow for fluctuations in seeing and cloud cover. Using the $\textit{wqed}$ software suite \citep{Thompson09}, we fit a second-order polynomial to the data to remove the long-term trend caused by atmospheric extinction, and applied a timing correction to each observation to account for the motion of the Earth around the barycenter of the solar system.

\begin{figure}[t]
\centering{\includegraphics[width=\columnwidth]{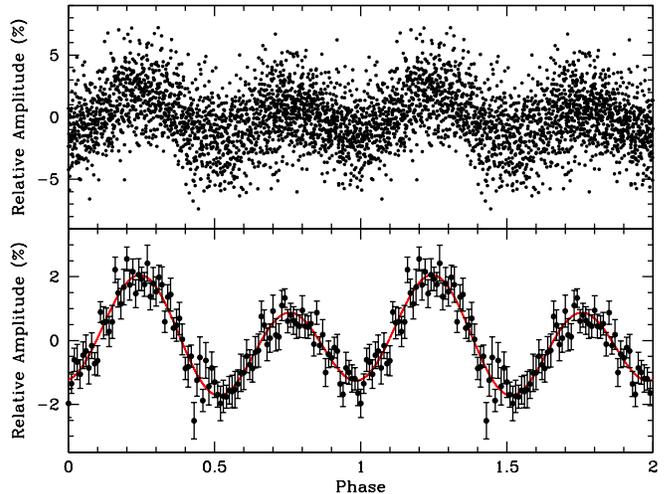}}
\caption{High-speed photometry of J1741 over 9.5 hr (top panel), folded at the orbital period, 87.998 minutes. The bottom panel shows the same light curve binned by 100 phase bins. The solid line includes 1.46\% amplitude ellipsoidal variations, a 0.59\% Doppler beaming signal, and a 0.23\% reflection effect. \label{fig:J1741lc}}
\end{figure}

Figure~\ref{fig:J1741lc} shows all $2{,}156$ Argos light curve points obtained for J1741 (top panel) folded over the best-fit orbital period from the radial velocity variations, 87.998 minutes. The bottom panel shows the same light curve binned into 100 orbital phase bins. The red solid line corresponds to the best fit orbital solution, as discussed in the following analysis.

\subsection{Analysis and Discussion}

There are four major effects that can cause photometric variability in the primary of a binary system: Doppler beaming, reprocessed light from the secondary (reflection), ellipsoidal variations, and eclipses. Here, and throughout, we will consider the ELM WD to be the primary, as it is the only visible component of this single-lined spectroscopic binary.

Given a high enough radial velocity variation, Doppler beaming (also referred to as Doppler boosting or relativistic beaming) will act to modulate stellar flux upon approach or recession (see, e.g., \citealt{Zucker07}). We treat this as a $\sin{\phi}$ modulation at the orbital period, where $\phi=0$ is defined by the spectroscopic conjunction (the point at which the primary is farthest away from us). Given our blue-bandpass filter, a $T_{\rm eff}=9{,}790\pm240$ K, and a radial velocity amplitude $K=508\pm4$ km s$^{-1}$, we expect a $0.58\pm0.02$\% Doppler beaming signal in J1741 \citep{Shporer10}. When present, this signal is usually evident as a strong asymmetry in the maxima of ellipsoidal variations.

Irradiation of the primary by the secondary can cause a reflection effect, which will also sample the orbital period, with a maximum at $\phi=0$. Thus, we treat the reflection effect as a $\cos{\phi}$ modulation at the orbital period.

Finally, tidal distortions of the primary will cause ellipsoidal variations. The dominant modulation occurs when the larger face comes into view twice per rotation, effectively a $\cos{2\phi}$ modulation of the spin period, which would be the orbital period for a synchronized system. However, since tidal distortions do not cause a perfectly ellipsoidal shape, we treat the ellipsoidal variations as harmonics to the first four $\cos{\phi}$ terms as derived in \citet{Morris93}.

To best characterize the variability in the light curve, we perform a Monte Carlo analysis. We create $10^5$ synthetic light curves by replacing the measured flux $f$ with $f + g~\delta f$, where $\delta f$ is the error in flux and $g$ is a Gaussian deviate with zero mean and unit variance. We then fit each light curve with a five-parameter model that includes an offset, and the (co)sine terms for Doppler beaming, ellipsoidal variations, reflection, and the first harmonic of the orbital period (e.g., \citealt{Shporer10,Sirko03,Mazeh10}). Our high-speed photometric observations reveal significant amplitudes to the $\sin{\phi}$, $\cos{\phi}$, and $\cos{2\phi}$ terms (the rest are consistent with zero within the errors; our full results can be found in Table~\ref{tab:par}).

Additionally, we have computed a Fourier transform of our time-series photometry. With less than 10 hr of data spread over four months, there is much alias structure around our highest peak, but the third-highest alias is identical to half the radial velocity period, within the formal error estimates: $P_{ELV}=2{,}639.920\pm0.016$ s. This peak occurs with $1.43\pm0.06$\% amplitude. (These are formal least-squares errors; if we did not have a guess for the period a priori, the spread in alias peaks would make a more realistic error, yielding $P_{ELV}=2{,}667\pm27$ s.) After pre-whitening by this first periodicity, the second-highest alias of the remaining peak occurs at $P_{DB}=5{,}280.27\pm0.13$ s ($0.67\pm0.06$\% amplitude).

However, since we have signals that may occur at $\cos{\phi}$ and $\sin{\phi}$, our Monte Carlo analysis yields a more robust estimate for the amplitude of the Doppler beaming effect: $0.59\pm0.06$\% amplitude. The Doppler beaming signal yields no new physical information about the binary, but since it matches the expected value of $0.58\pm0.02$\% within our errors, it confirms the radial velocity amplitude observed in this system, as well as the spectroscopically determined temperature for the ELM WD. This is also by far the strongest measurement of Doppler beaming in a binary system from the ground, even greater than the 12.75-minute double WD binary J0651 \citep{BrownJ0651}, due to its cooler effective temperature \citep[see Equations (2) and (3) in][]{Shporer10}.

The strongest modulation measured in our Monte Carlo analysis occurs for the $\cos{2\phi}$ term, consistent with ellipsoidal variations, with an amplitude of $1.46\pm0.06$\%. These variations are extremely useful for constraining the radius of the tidally distorted ELM WD. To do so, we use the discrete Fourier series for ellipsoidal variations from \citet{Morris93}. Equation (1) in that work yields a theoretical prediction for the ellipsoidal variation amplitude dominated by
$$ L(\phi) / L_0 = \frac{-3 (15 + u_1) (1 + \tau_1) (r_1 / a)^3 q \sin^2 i}{20 (3-u_1)} \cos(2\phi)  $$
based on the mass function $q=m_2/m_1$, the limb-darkening ($u_1$) and gravity-darkening ($\tau_1$) coefficients for the primary, the semimajor axis of the system $a$, the orbital inclination $i$, and the radius of the primary $r_1$. Using reasonable values for the limb-darkening and gravity-darkening coefficients, we constrain the radius of the primary as a function of the orbital inclination of the system. Here we assume a linear limb darkening law and set $u_1=0.36$, but consider values between $u_1=0.1-0.5$, given the observations of \citet{Parsons11}. Since we are most interested in setting a lower limit on the radius of the ELM WD, we also assume that the stellar flux is purely radiative, and set $\tau_1=1.0$. This assumption may be well justified, as all models for 10,000 K low-mass WDs down to 0.17 \msun\ have no surface convection (J. Panei 2011, private communication).

\begin{figure}[t]
\centering{\includegraphics[width=\columnwidth]{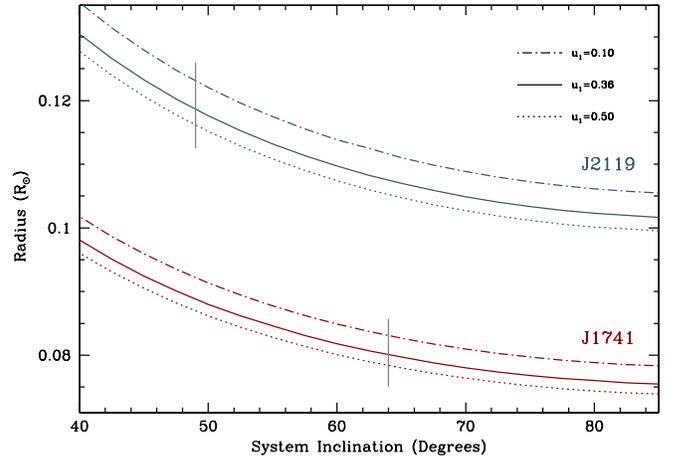}}
\caption{Allowed values for the radius, modulo the system inclination, to explain the 1.46\% amplitude ellipsoidal variations in J1741 and the 1.56\% amplitude ellipsoidal variations in J2119. Our estimates include different values of the limb-darkening coefficient, but we assume a purely radiative WD atmosphere ($\tau_1=1.0$). For the 0.16 \msun\ ELM WD in J1741, its companion would be a neutron star if the inclination is less than about 64$\arcdeg$, as marked by the vertical line, and if the inclination is less than 41$\arcdeg$, the companion would exceed 3.2 \msun. For the 0.17 \msun\ ELM WD in J2119, its companion would be a neutron star if the inclination is less than about 49$\arcdeg$. \label{fig:inc_radius}}
\end{figure}

Figure~\ref{fig:inc_radius} shows the radius of J1741 as a function of the unknown inclination angle. Here we calculate the radius using three different values of the limb-darkening coefficient. Our photometry rules out 2\% or deeper primary eclipses, which constrains the system inclination to be less than 83$\arcdeg$. The minimum radius of the ELM WD in J1741 is thus 0.074 \rsun. This also implies that the mass of the unseen companion exceeds 1.11 \msun.

The temperature and surface gravity estimates for J1741 do not overlap with the theoretical models of \citet{Panei07}, making our radius measurement more important. \citet{Panei07} predict a radius of 0.09 \rsun\ for a 1 Myr old 0.16 \msun\ WD with $T_{\rm eff}=9{,}200$ K and $\log{g}=5.7$ cm s$^{-2}$. Our radius measurement for J1741 provides the first observational constraints on such a low surface gravity WD.

Using the values that best reproduce the $\cos{2\phi}$ term for ellipsoidal variations from \citet{Morris93}, the expected amplitude for the $\cos{\phi}$ term is less than 0.02\%. This is not compatible with our observations: we measure $0.23\pm0.06$\% variations in the $\cos{\phi}$ term, which is a 3.8$\sigma$ detection.

This variation indicates that the primary is brighter by 0.23\% when its heated surface is facing us at $\phi = 0$. Assuming that the unseen secondary is a 1.2 \msun C/O WD with a typical radius of 0.0087 \rsun\ ($i=75\arcdeg$), the secondary needs to be hotter than 78,000 K to reproduce a 0.23\% reflection effect \citep{Kopal59}. Such a companion would easily be detected in the SDSS photometry and the MMT spectroscopy data. Hence, the companion may be a pulsar.

Millisecond pulsars are known to ``zap'' their companions, e.g., the black widow pulsar system PSR 1957+20 \citep{Fruchter88} and the double pulsar system J0737-3039 \citep{Burgay03,Lyutikov05}. This irradiation, which can lead to ablation of the companion star, is believed to be caused by relativistic particles emitted by the pulsar. The observed reflection effect suggests a pulsar companion, which would require $i\le 65\arcdeg$. If the inclination of the system is roughly 60$\arcdeg$, the observed ellipsoidal variations would constrain the radius to $\ge0.078$ \rsun. A pulsar companion is unlikely to have a mass exceeding 3.2 \msun, putting an upper limit on the ELM WD's radius of 0.102 \rsun\ (if $u_1=0.1$). Although not suggested in the models, we may also consider the scenario where some of the ELM WD surface is convective ($\tau_1=0.48$), in which case gravity darkening will increase the maximum allowable radius to 0.112 \rsun\ if $m_2=3.2$ \msun.



\section{SDSS J211921.96-001825.8}

The ELM WD J2119 has $T_{\rm eff}=10{,}360 \pm 230$ K, $\log{g}=5.36\pm0.07$, and $M\sim0.16$ \msun\ \citep{BrownELMi}. It is a binary with a $2.0825\pm0.0010$ hr orbital period and a mass function of $0.501\pm0.016$ \msun, yielding a minimum mass for the companion of 0.75 \msun.

\subsection{The Light Curve}

\begin{figure}[t]
\centering{\includegraphics[width=\columnwidth]{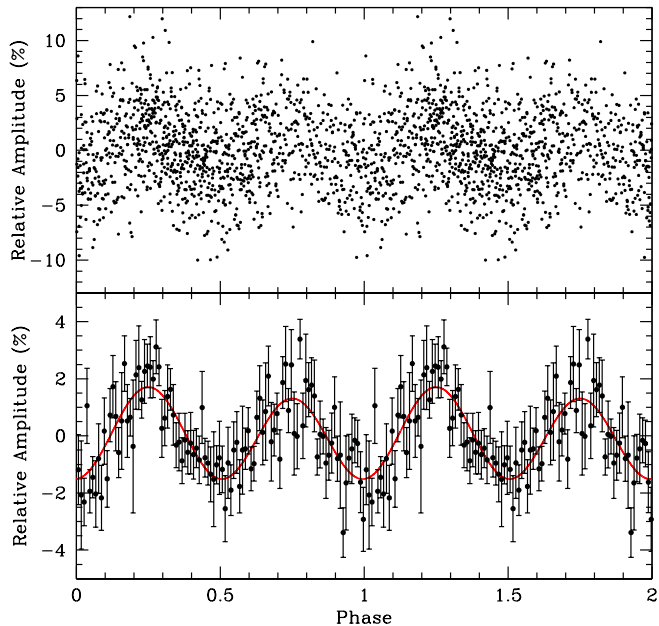}}
\caption{High-speed photometry of J2119 over 10.4 hr (top panel), folded at the orbital period, 124.95 min. The bottom panel shows the same light curve binned by 100 phase bins. The solid line shows a 1.51\% amplitude signal at half the orbital period plus a 0.20\% Doppler beaming effect at the orbital period. \label{fig:J2119lc}}
\end{figure}

Observations of J2119 were carried out at the McDonald Observatory over three nights in 2011 July, for a total of $1{,}252$ points over more than 10 hr. All observations of this $g=20.0$ mag WD were with 30 s exposures, and were reduced in an identical manner as those of J1741.

Figure~\ref{fig:J2119lc} shows the Argos light curve of J2119 (top panel) folded over the best-fit orbital period from the radial velocity variations, 124.95 minutes. The bottom panel shows the same light curve binned into 100 orbital phase bins. The red solid line shows our best-fit model, as discussed below.

\subsection{Analysis and Discussion}

We model the light curve of J2119 using a five parameter model as in J1741. Given the estimate for J2119 of $T_{\rm eff}=10{,}360\pm230$ K and a radial velocity amplitude $K=383\pm4$ km s$^{-1}$, we expect a $0.41\pm0.02$\% Doppler beaming signal. Only the $\cos{2\phi}$ term for ellipsoidal variations and the $\sin{\phi}$ term for Doppler beaming appear with any statistical significance in the Monte Carlo analysis of the Argos light curve (our full results can be found in Table~\ref{tab:par}). This analysis yields an estimate for the observed amplitude of the ellipsoidal variations at $1.56\pm0.13$\% and that of Doppler beaming signal at $0.20\pm0.12$\%. Argos photometry rules out 3\% or deeper primary eclipses, which constrains the system inclination to be less than 85$\arcdeg$.

Again, we have computed a Fourier transform of our time-series photometry as an additional check. The highest peak in our dataset occurs at $3{,}745.9\pm2.7$ s ($1.47\pm0.14$\% amplitude), at half the orbital period. After pre-whitening by this peak, the second-highest alias of the remaining peak occurs at $7{,}592\pm25$ s ($0.65\pm0.14$\% amplitude).

The disparity between amplitudes for the Doppler beaming signal in our Monte Carlo and Fourier analysis is notable, but considering that our photometric data cover less than five orbits for this relatively faint system, we will not over-interpret our results for this signal. We require more data to use the signal as a constraint on the radial velocity amplitude and spectroscopic temperature.

On the other hand, the ellipsoidal variations are detected with high significance in both analyses. Again, we use the ellipsoidal variations to constrain the radius modulo the inclination of the system. We use the same limb- and gravity-darkening coefficients as in J1741, and our constraints on the radius of the ELM WD in the J2119 system are presented in Figure~\ref{fig:inc_radius}. Assuming that the ELM WD is purely radiative, we find a minimum radius of the ELM WD primary to be 0.10 \rsun. This radius is slightly larger than the radius estimate for J1741. J2119 is estimated to be 600 K warmer than J1741. Even though this is only a relatively small temperature difference, 0.16\msun\ WDs are predicted to cool down only by 1900 K after 10 Gyr \citep{Panei07} and decrease in size by half. Hence, the relatively small temperature difference between J2119 and J1741 may be enough to explain the difference in their inferred radii.



\section{Conclusions}

\begin{figure*}
\centering
\includegraphics[width=0.93\textwidth]{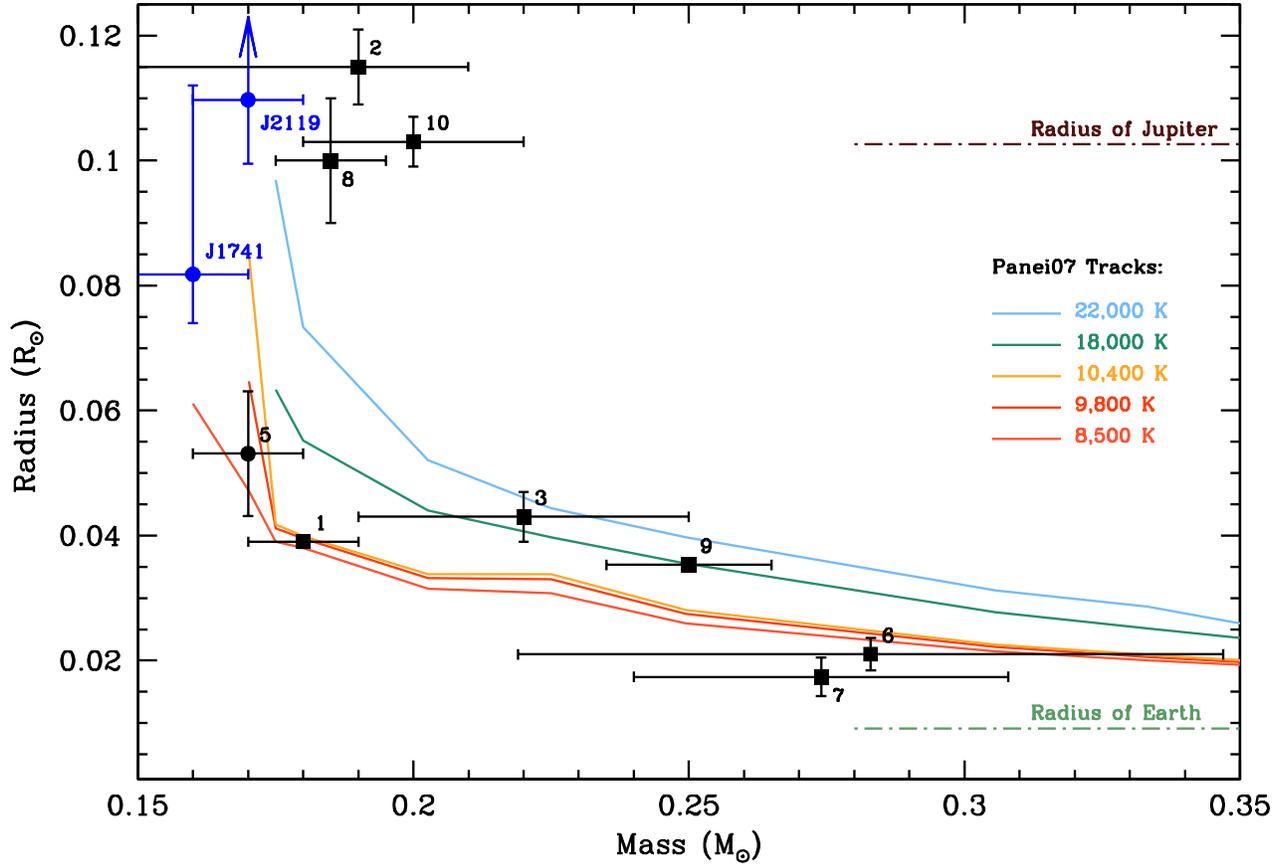}
\caption{Observed mass and radius determinations for low-mass (He-core) WDs. The radius measurements for points marked by squares came from eclipsing systems and the dots from the amplitude of ellipsoidal variations, as in this work. The points for J1741 and J2119 represent the case where $i=60\arcdeg$. The mass-radius tracks from \citet{Panei07} cover a range of temperatures, ranging from nearly 22,000 K, as seen in CSS 41177A, to nearly 8,500 K, as seen in NLTT 11748. Our mass/radius values come from several references: (1) NLTT 11748 \citep{Steinfadt10a,Kilic10}; (2) KOI81B \citep{vankerkwijk10,Rowe10}; (3) KOI74B \citep{vankerkwijk10}; (4, not pictured; $0.26\pm0.04$ \msun, $0.15\pm0.01$ \rsun) KHWD3 \citep{Carter11}; (5) J0106 \citep{KilicJ0106}; (6 and 7) CSS 41177 A and B \citep{Parsons11}; (8) GALEX J1717 \citep{Vennes11}; (9) J0651 \citep{BrownJ0651}; (10) KHWD4 \citep{Breton11}; and J1741 and J2119 (this work). We have included the radii of Earth and Jupiter for reference. \label{fig:massradius}}
\end{figure*}

We present the discovery of two new tidally distorted ELM WDs. We use the amplitude of the observed ellipsoidal variations to constrain the radius of this rare, fluffy class of WDs. J1741 and J2119 are larger than 0.074 and 0.10 \rsun, respectively. These two objects plus {\em Kepler} Hot White Dwarf 3 \citep[$R\sim0.15$\rsun][]{Carter11} and GALEX J171708.5+675712 \citep[$R\sim0.1$\rsun][]{Vennes11} are the largest radii WDs ever observed. Unlike the typical Earth-sized 0.6 \msun\ C/O-core WDs, these $\sim$0.2 \msun\ He-core WDs are similar in size to a giant planet like Jupiter.

We have detected a Doppler beaming signal in the photometry of both systems, as a result of the high-amplitude radial-velocity variations. The effect measured in J1741, with $0.59\pm0.06$\% amplitude, represents the strongest Doppler beaming signal detected so far in a binary system.

J1741 remains an intriguing object. It is so far the only tidally distorted WD that also shows a small but significant reflection effect. We observe a $0.23\pm0.06$\% flux modulation at the orbital period of this ELM WD which peaks at $\phi=0$. No comparable signal is observed in J2119, which shows even higher-amplitude ellipsoidal variations than J1741. Follow-up photometry at a larger telescope is required to confirm this reflection effect. If real, this may be due to the irradiation of the WD by a pulsar companion that may be detected in the radio or X-ray observations. We are pursuing follow-up X-ray observations of J1741 with the {\em XMM-Newton} telescope. If the companion is a neutron star, the inclination of J1741 is less than 65$\arcdeg$ and the WD has a radius $\ge0.078$ \rsun.

We measure ellipsoidal variations in J1741 and J2119 of $1.46\pm0.06$\% and $1.56\pm0.13$\% amplitude, respectively. A Fourier analysis of these high-amplitude variations do not indicate a detectable difference between the period of the radial velocity variations and twice the period of the ellipsoidal variations (see Table~\ref{tab:par}), although there is too much aliasing in these observations to say with certainty that the system is synchronized, such that the ELM WD rotation period matches the orbital period found from the radial velocity variations.

Using these ellipsoidal variations, we have put limits on the radii of two of the lowest-mass WDs ever directly observed. There are less than a dozen other empirical mass-radius determinations for low-mass (He-core) WDs, and we plot our results with the theoretical mass-radius relations from evolutionary models by \citet{Panei07} in Figure~\ref{fig:massradius}.

Evolution models predict that He-core WDs with masses less than 0.17 \msun\ should sustain sufficient hydrogen shell burning \citep{Serenelli02,Panei07,Steinfadt10b}. In fact, a majority of the flux from these $\leq0.17$ \msun\ WDs comes from this residual burning. This should in turn cause a considerable increase in their radius, which has so far been borne out in our empirical mass-radius determinations.

\begin{deluxetable*}{lclc}
\tablewidth{350pt}
\tablecaption{Physical parameters\label{tab:par}}
\tablehead{\multicolumn{4}{c}{\bf{J1741}}}
\startdata
\\[-0.7em]
RA (J2000)    & 17:41:40.49    & Dec (J2000)            & +65:26:38.7      \\
\teff\ (K)    & $9{,}790\pm240$ & \logg\ (cm s$^{-2}$)  & $5.19\pm0.06$    \\
$m_1$ (\msun) & $0.16\pm0.01$  & $g_0$ (mag)            & $18.271\pm0.022$ \\
$P_{orb}$ (s) & $5{,}279.90\pm0.86$ & $P_{ELV}$ (s)      & $2{,}639.920\pm0.016$ \\
$\cos(\phi)_{\rm{amp}}$ (\%) & $0.23\pm0.06$ & $\cos(2\phi)_{\rm{amp}}$ (\%) & $1.46\pm0.06$ \\
$\sin(\phi)_{\rm{amp}}$ (\%) & $0.59\pm0.06$ & $\sin(2\phi)_{\rm{amp}}$ (\%) & $0.01\pm0.06$ \\
$r_1$ (\rsun) & 0.074 --- 0.112 & $T_{0,ELV}$ (BJD$_{\mathrm{TDB}}$) & 2455686.76251(21) \\
\cutinhead{\bf J2119}
\\[-0.7em]
RA (J2000)    & 21:19:21.96    & Dec (J2000)            & $-$00:18:25.8    \\
\teff\ (K)    & $10{,}360\pm230$ & \logg\ (cm s$^{-2}$) & $5.36\pm0.07$    \\
$m_1$ (\msun) & $0.17\pm0.01$  & $g_0$ (mag)            & $20.000\pm0.021$ \\
$P_{orb}$ (s) & $7{,}496.9\pm3.5$ & $P_{ELV}$ (s)        & $3{,}745.9\pm2.7$  \\
$\cos(\phi)_{\rm{amp}}$ (\%) & $0.02\pm0.13$ & $\cos(2\phi)_{\rm{amp}}$  (\%) & $1.56\pm0.13$ \\
$\sin(\phi)_{\rm{amp}}$ (\%) & $0.20\pm0.13$ & $\sin(2\phi)_{\rm{amp}}$  (\%) & $0.06\pm0.13$ \\
$r_1$ (\rsun) & 0.10 --- 0.13 & $T_{0,ELV}$ (BJD$_{\mathrm{TDB}}$) & 2455769.84065(66) \\[-0.7em]
\enddata
\end{deluxetable*}

The orbital periods in both J1741 and J2119 are decaying due to the emission of gravitational radiation. It is possible to use the time-of-minimum of the ellipsoidal variations in both objects to detect this orbital period decay. This rate of the orbital period change is sensitive to the mass of the unseen secondary, but for J1741, the orbital period is changing by at least $-3.6 \times 10^{-13}$ s s$^{-1}$. This would cause the observed minimum to occur more than 8 s sooner within 15 years. With less than 10 hr of data spread over four months, the formal error on the observed phase is roughly 20 s, which will make this measurement difficult in less than a decade. If the companion is indeed a pulsar, the orbital period would change by at least $-4.5 \times 10^{-13}$ s s$^{-1}$, and the timing measurements of the pulsar could make this measurement possible in just a few years. For J2119, the orbital period is changing by at least $-3.4 \times 10^{-13}$ s s$^{-1}$. We have included our $T_0$ ephemeris measurements for this first epoch of observations in Table~\ref{tab:par}.


\acknowledgments

The observations of J2119 in 2011 July were assisted by G. Miller, S. Wang, G. Earle, J. Pelletier, and A. Rost, undergraduate students at the University of Texas Freshmen Research Initiative. J.J.H., M.H.M. and D.E.W. gratefully acknowledge the support of the NSF under grant AST-0909107 and the Norman Hackerman Advanced Research Program under grant 003658-0252-2009.

\end{document}